\documentclass[12pt]{article}
\usepackage{amsfonts, amscd}
\usepackage{amssymb}
\usepackage{eucal,eufrak}
\usepackage{verbatim}
\usepackage{latexsym}
\usepackage{epsfig}

\def\IH{{\mathbb H}}
\def\IR{{\mathbb R}}
\def\IP{{\mathbb P}}
\def\IC{{\mathbb C}}

\font\teneufm=eufm10
\font\seveneufm=eufm7
\font\fiveeufm=eufm5
\newfam\eufmfam
\textfont\eufmfam=\teneufm
\scriptfont\eufmfam=\seveneufm
\scriptscriptfont\eufmfam=\fiveeufm

\newcommand{\infinity}{\ensuremath{\infty}}
\newcommand{\dd}{\ensuremath{\partial}}

\newcommand{\fdot}{\partial_t f}

\newcommand{\fddot}{\partial^2_t f}
\newcommand{\drf}{\partial_r f}
\newcommand{\drrf}{\partial^2_r f}

\newcommand{\delr}{\triangle r}
\newcommand{\delt}{\triangle t}

\newcommand{\grad}{\raisebox{.5 ex}{\ensuremath{\bigtriangledown}}}

\begin{document}

\title{Fast and Slow Blowup in the $S^2$ Sigma Model and
(4+1)-Dimensional Yang-Mills Model\thanks{This work was partially
supported by the Texas Advanced Research Program, and carried out at
the University of Texas.}}

\author{Jean Marie Linhart\thanks{Applied Science Fiction; 
8920 Business Park Drive; Austin,
TX 78759; \hfill\break jlinhart@asf.com.}
\and 
Lorenzo A. Sadun\thanks{
Department of Mathematics; University of Texas at Austin; Austin, TX
78712;  \hfill\break sadun@math.utexas.edu}
} 

\date{\today}

\maketitle

\begin{abstract}
We study singularity formation in spherically symmetric solutions of
the charge-one and charge-two sector of the (2+1)-dimensional $S^2$ 
$\sigma$-model and the (4+1)-dimensional Yang-Mills model, near the adiabatic limit.  
These equations
are non-integrable, and so studies are performed numerically on
rotationally symmetric solutions using an iterative finite differencing
scheme that is numerically stable.  
We evaluate the accuracy of predictions made with the
geodesic approximation. We find that the geodesic approximation is extremely
accurate for the charge-two $\sigma$-model and the Yang-Mills model, both
of which exhibit fast blowup.  The charge-one $\sigma$-model exhibits slow
blowup.  There the geodesic approximation must be modified by applying
an infrared cutoff that depends on initial conditions.
\end{abstract}

Mathematics Subject Classification: 35-04, 35L15, 35L70, 35Q51, 35Q60

Physics and Astronomy Classification: 02.30.Jr, 02.60.Cb

\newpage

\section{Introduction, Background and Results}

The geodesic approximation \cite{Manton} 
is a popular tool for theorizing on the
dynamics of systems with known spaces of stationary solutions.  As the
velocity tends toward zero, evolution should occur close to geodesics
on the space of the stationary solutions.

But do equations in fact evolve as the geodesic approximation
predicts?

We study three systems to which the geodesic approximation has been
applied.  We study the charge-one sector of the (2+1)-dimensional
$S^2$ $\sigma$-model (hereafter referred to as the ``charge-one 
$\sigma$-model''), the charge-two sector of the (2+1)-dimensional $S^2$ 
$\sigma$-model (the ``charge-two $\sigma$-model''), and the (4+1)-dimensional
Yang-Mills model.  The $S^2$ $\sigma$-model is sometimes called the
$O(3)$ $\sigma$-model or the $\IC \IP^1$ model, and the localized solutions are
called ``lumps''.  In all cases the geodesic approximation suggests
that localized solutions should shrink to zero size (``blow up'') in
finite time.

Blowup phenomena can be categorized as ``fast'' or ``slow''.  In slow
blowup, all relevant speeds go to zero as the singularity is
approached.  In fast blowup the relevant speeds do not go to zero.
Shatah and Struwe \cite{SS} proved
that there cannot be spherically symmetric fast blowup in the charge
one $\sigma$-model.  In particular, the approach to blowup cannot be
asymptotically self-similar, as suggested by the geodesic
approximation.  This theorem does not apply to the charge-two 
$\sigma$-model, or to the (4+1)-dimensional Yang-Mills system. 

In each case, we model the equations in a simple and robust numerical
scheme for which the stability can be verified.  Then we know the
numerical scheme does not add features to the evolution of the
differential equation.  Consequently, we can make meaningful comparisons of
the predictions of the geodesic approximation with the actual
evolution of these equations.  We observe slow blowup in the charge
one $\sigma$-model, and fast blowup in the charge-two $\sigma$-model and the
Yang-Mills model, in accordance with the Shatah-Struwe theorem.

This is not the first paper to numerically observe blowup in these
models.  However, our intent is not merely to confirm blowup.  We
quantify the deviations of the true trajectories from the geodesic
approximation, in particular for the charge-one $\sigma$-model, where these
deviations are substantial.  We also suggest a modification of the
geodesic approximation for the charge-one $\sigma$-model, applying a
dynamically generated cutoff to certain integrals to obtain
dramatically improved accuracy.  In the charge-two $\sigma$-model and the
Yang-Mills model, we find that the geodesic approximation is extremely
accurate, and no such modification is needed.

The literature on $\IC \IP^1$ lump dynamics is extensive, especially in the
charge-one sector, and begins with Ward's suggestion \cite{Ward} of
applying Monton's geodesic approximation \cite{Manton}
to this problem.  The dynamics of
lump scattering, in this approximation, were studied in depth by Leese
\cite{Leese} and Zakrzewski \cite{Zakr} on $\IR^2$, and by Speight on
$S^2$ \cite{Speight1} and on other Riemann surfaces \cite{Speight2}.

In 1990, Leese, Peyrard and Zakrzewski \cite{LPZ} studied the stabilty
of numeric solitons, using a square grid and the traditional
Runge-Kutta numerical approach, and found that lumps on a discretized
system are inherently subject to shrinkage.  This is of direct
physical interest, since many physical systems (e.g., 
an array of Heisenberg ferromagnets) are naturally discretized.  
However, it raises the questions
of whether a different discretization might avoid introducing
spontaneous shrinking.  In this paper we exhibit one that does.

The results of \cite{LPZ} also demonstrate the importance of doing a
careful stability analysis of whatever numerical scheme one applies,
to separate properties of the underlying differential equations from
phenomena introduced by the discretization.  The numerical methods of this
paper were chosen to make such analysis possible \cite{Linhart}.

The most important numerical test of the geodesic approximation of
blowup, for the rotationally symmetric charge-one $\sigma$-model, was
done by Piette and Zakrzewski \cite{PZ}.  In the charge-one sector, the
Lagrangian for lump dynamics in the geodesic approximation has an
infrared log divergence.  To remedy this, Piette and Zakrzewski cut off
their integrals at a large distance $R$, computed trajectories, and
then took the $R \to \infty$ limit.  This procedure predicts fast
blowup ($a(t) \sim t_0-t$, where $a(t)$ is the size of the lump at time
$t$), contradicting the Shatah-Struwe theorem.  This prediction is also in
mild contradiction of Piette and Zakrzewski's numerical results.  They
found that $a(t)$ shrank slightly slower than linear, which they modeled
as a power law with exponent slightly greater than one.  

More recently, Bizo\'n, Chmaj and Tabor studied the universality of
blowup in the charge-one $\sigma$-model \cite{BCT}.  They provide
numerical evidence that the existence of blowup does not depend on the
initial conditions (as long as there is enough energy), that the shape
of the lump prior to blowup is universal, that the rate of blowup is slow
($da/dt \to 0$ as $t \to t_0$) but that the rate is dependent on initial 
conditions.

In this paper, we extend the results of \cite{PZ} in several ways.
First, we obtain more accurate numerical data on the rate of blowup.
We see that blowup in the rotationally symmetric charge-one 
$\sigma$-model deviates from linear by a log correction, not a power law.
Second, we show that this can be understood within the framework of
the geodesic approximation.  The cutoff Lagrangian of \cite{PZ} gives
the correct dynamics if we do {\em not} take the $R \to \infty$ limit.
Rather, $R$ should be understood as a dynamically generated cutoff
that depends on the initial conditions.  ($R$ is of order
$|a(0)/a'(0)|$).  The resulting trajectories are qualitiatively
similar to those found by Speight for lump dynamics on a sphere
\cite{Speight1}.  The difference is that Speight's cutoff (the radius
of the sphere) is independent of initial conditions, while ours is
fundamentally dynamic.

Third, we show how the shape of the lump depends on the speed of its
shrinkage. In the geodesic approximation, the lump at all times looks
like a static solution to the equations.  We show that this is only
approximately correct, and we quantify the deviations.  In particular,
we see that, shortly before blowup, lumps have the property that a small
disk around the origin maps onto all of the target $S^2$.

Finally, we model the charge-two $\sigma$-model and the Yang-Mills model
as a comparison to the charge-one $\sigma$-model.  The form of the
equations, the method of discretization, and the numerical stability
analysis are similar for all three models.  The conventional wisdom is
that the Yang-Mills model should behave much like the charge-two
$\sigma$-model, and this is borne out by our numerics.
Moreover, in both these cases the rate of blowup
is extremely close to that predicted by the geodesic approximation.
This gives a standard of accuracy by which to judge the charge-one
$\sigma$-model.  To wit, the deviations from prediction on the charge-one
$\sigma$-model are much, much larger and are not due to numerical error.

This paper is organized as follows.  The $S^2$ $\sigma$-model and the
Yang-Mills model are reviewed in Section 2.  In Section 3 we present
our method for numerically integrating the PDEs, and discuss the
stability and accuracy of this method.  Our results appear in Section
4, first for the charge-one $\sigma$-model, then for the charge-two
$\sigma$-model, and finally for the Yang-Mills model.

\newpage

\section{Equations being modeled}

\subsection{The 2-dimensional $S^2$ $\sigma$-Model}

The two-dimensional $S^2$ $\sigma$-model has been studied extensively
over the past few years in \cite{Leese}, \cite{PZ},
\cite{LPZ},\cite{Speight1}, \cite{Speight2}, 
\cite{Ward}, \cite{Zakr} and \cite{Ioa}.  

It is a good toy model for studying two-dimensional analogues of
elementary particles in the framework of classical field theory.
Elementary particles are described by classical extended solutions of
this model, called lumps.  Such lumps can be written down explicitly
in terms of rational functions.  This model is extended to (2+1)
dimensions.  The previous lumps are static or time-independent
solutions, and, with the added dimension, the dynamics of these lumps
are studied.  The time dependent solutions cannot be constructed
explicitly, so in (2+1)-dimensions studies are performed numerically,
and analytic estimates are made by cutting off the model outside a
radius $R$.

We will see that the $S^2$ $\sigma$-model displays both slow blowup
and fast blowup.  The charge-one $\sigma$-model exhibits logarithmic
slow blowup, whereas the charge-two $\sigma$-model and the similar
Yang-Mills (4+1)-dimensional model exhibit fast blowup.

Identifying $S^2 = \IC \IP^1 = \IC \cup \{\infinity\}$ we can write
the Lagrangian density in terms of a complex scalar field $u$:
\begin{equation} L = \int_{\IR^2} \frac{|\partial_t{u}|^2}{(1 + |u|^2)^2} - 
\frac{|\grad u|^2}{(1 + |u|^2)^2}.\label{cp1lag}\end{equation}

The calculus of variations on this Lagrangian in conjunction with
integration by parts yields the following equation of motion for the
$\IC \IP^1$ model:
\begin{equation}
(1 + |u|^2)(\dd_t^2u - \dd_x^2 u - \dd_y^2 u) = 2\bar{u}((\dd_t u)^2 - (\dd_x u)^2
- (\dd_y u)^2) \label{genPDEb}\end{equation}
Here  $\bar{u}$ represents the complex conjugate of $u$.

We will look at the evolution of these equations
near the space of static solutions.  The static solutions are outlined in
\cite{Ward} among others.  The entire space of static solutions can be
broken into finite dimensional manifolds $\mathcal{M}_n$ consisting of
the harmonic maps of degree $n$.  If $n$ is a positive integer, then
$\mathcal{M}_n$ consists of the set of all rational functions of $z =
x + iy$ of degree $n$.  We restrict our attention to
$\mathcal{M}_1$ and $\mathcal{M}_2$, the charge-one sector and the 
charge-two sector. On these two sectors the static solutions (with $u(\infty)$
finite) all have the form
\begin{equation} u = \alpha + \frac{\beta}{z + \gamma}\label{genc1soln}\end{equation}
and 
\begin{equation} u  = \alpha + 
\frac{\beta z + \gamma}{z^2 + \delta z + \epsilon},
\label{genc2soln}\end{equation}
respectively, depending on up to five complex parameters $\alpha,
\beta, \gamma, \delta, \epsilon$.  To simplify further, we impose
rotational symmetry on the solution, which is respected by
(\ref{genPDEb}).  For this regime the (2+1)-dimensional model reduces
to a (1+1) dimensional model.  These solutions become
\begin{equation}
\frac{\beta}{z}\ \mbox{(on\ }\mathcal{M}_1\mbox{)}
\ \  {\rm and}\ \  \frac{\gamma}{z^2}\ \mbox{(on\ }
\mathcal{M}_2\mbox{),}\end{equation} 
with $\beta$ and $\gamma$ real.  Note that by applying rotational symmetry
we have also forced $u(\infty)$ to be zero.\footnote{There are also 
rotationally symmetric static solutions with $u(\infty)=\infty$.  Since
the equations of motion are invariant under the transformation $u \to 1/u$,
these behave identically to the solutions discussed here, and do not have to 
be considered separately.} 
The geodesic approximation predicts that
solutions evolve close to \begin{equation} \frac{\beta(t)}{z}\ \  {\rm or}\ \ \frac{\gamma(t)}{z^2}.\end{equation}

This paper is about the actual evolution of these equations.   
Instead of looking at $\beta(t)$ or $\gamma(t)$, 
we look for the most general rotationally
symmetric solution to (\ref{genPDEb}).  We find the evolution of: 
\begin{equation} \frac{f(r,t)}{z}\ \ \mbox{or} \ \ \frac{f(r,t)}{z^2}.\end{equation}
The difference between the actual evolution of $f(0,t)$ and that predicted
in the geodesic approximation, and the difference between 
the profile  $f(r,t)$ (with $t$
fixed) from $\beta(t)$ (which is constant with respect to $r$)  
gives us a means to gauge how accurately the
evolution of this equation can be modeled by the geodesic approximation.

It is straightforward to calculate the evolution equation for $f(r,t)$.  
For $u=\frac{\textstyle{f(r,t)}}{\textstyle{z}}$ it is:
\begin{equation} \fddot = \drrf + \frac{3\drf}{r} - \frac{4r\drf}{f^2 + r^2} + 
\frac{2f}{f^2 + r^2}\left((\fdot)^2 - (\drf)^2\right).\label{PDEb}\end{equation}
For $u=\frac{\textstyle{f(r,t)}}{\textstyle{z^2}}$ it is:
\begin{equation} \fddot = \drrf + \frac{5\drf }{r} - \frac{8r^3\drf }{f^2 + r^4} +
\frac{2f}{f^2 + r^4}\left((\fdot)^2 -
(\drf)^2\right).\label{PDEc}\end{equation} 

In both cases the static solutions are $f(r,t) =$ constant.  
In the charge-one sector this constant is the length scale, while
in the charge-two sector it is the square of the length scale. 
In the geodesic approximation, $f(r,t)$ depends on $t$ but not on $r$.
Blowup occurs when the length scale becomes zero, that is $f(0,t)=0$. 
Even apart from the geodesic approximation, a singularity appears when
$f(0,t)=0$. This is what we call the instant of blowup.

Note that there is nothing singular about $f(r,t)$ equaling zero when
$r \ne 0$.  That merely indicates that $u$ maps the entire 
circle $z=re^{i\theta}$ to the south pole of $S^2$.  We shall see that
this always happens,
for smaller and smaller values of $r$, shortly before blowup.  

\subsection{The Yang-Mills Model}

The Yang-Mills Lagrangian in 4 dimensions is a generalization of
Maxwell's equations in a vacuum, and is discussed at length in
\cite{Atiyah}.  We can regard this problem as being that of a motion
of a particle, where we wish our particles to have certain internal
and external symmmetries, which give rise to the various geometrical
objects in the problem.  The states of our particles are given by
gauge potentials or connections, denoted $A$, on $\IR^4$, and we identify
$\IR^4$ with the quarternions $\IH$: $\vec{x} = x_1 + x_2 i + x_3 j +
x_4 k$.  The gauge potentials have values in the Lie algebra of
$SU(2)$ which can be viewed as purely imaginary quarternions.  
The curvature $F_{ij} = \dd_iA_j - \dd_jA_i +
[A_i,A_j]$, where $[A_i,A_j] = A_iA_j - A_jA_i$ is the bracket in the
Lie Algebra, gives rise to the potential $V(A) = |F_{ij}|^2$
which is a nonlinear function of $A$.  The static action is:
\begin{equation} S = \frac{1}{2}
\int \ |F_{ij}|^2 d^4\vec{x}. \label{YMPLagr}
\end{equation}
The space of finite-action configurations breaks up into topological sectors,
indexed by the instanton number.  We are interested in the sector with
instanton number one.
The local minima of (\ref{YMPLagr}) are the instantons on 4
dimensional space.

We now introduce a time variable $t$ and 
consider the wave equation generated
by this potential with Lagrangian:
\begin{equation} L = \frac{1}{2}\int |\partial_t{A_i}|^2
-\frac{1}{2} |F_{ij}|^2 d^4\vec{x}. \label{YMLagr}
\end{equation}
Via the calculus of variations, the evolution equation for this
dynamical model is\footnote{Note that we have defined
$A=(A_1,A_2,A_3,A_4)$ to be a time-dependent connection on $\IR^4$,
not a connection on $\IR^{4+1}$.  
However, the dynamics of Yang-Mills connections on $\IR^{4+1}$ is
almost identical.  In that model, $A=(A_0, A_i)$ and the $|\partial_t
A_i|^2$ term in the Lagrangian (\ref{YMLagr}) 
is replaced by $|F_{0i}|^2$.  After
obtaining the equations of motion we apply the gauge choice
$A_0=0$. The equations of motion then reduce to (\ref{genPDEa}), plus
a Gauss' Law constraint $\partial_t(\partial_i A_i)=0$ that comes from
varying the Lagrangian with respect to $A_0$.  This constraint is identically
satisfied by the ansatz (\ref{YMansatz}).  Our results therefore apply to
this model as well as to time-dependent connections on $\IR^4$.}
\begin{equation} \partial_t^2{A_i} = 
-\grad_j F_{ij}.\label{genPDEa}\end{equation}

Theoretical work on the validity of the geodesic 
approximation for an analogous problem, 
the monopole solutions to the Yang-Mills-Higgs theory
on $(3+1)$-dimensional Minkowski space, is presented in \cite{Stuart}.

The static solutions to equation (\ref{genPDEa}) are simply the 4
dimensional instantons investigated in \cite{Atiyah}.  All degree one 
instantons take the form:
\begin{equation} A(x) = \frac{1}{2}\left\{\frac{(\bar{x} - \bar{a})dx - d\bar{x} (x -a)}{\lambda^2 + |x-a|^2}\right\}
\qquad x = x_1 + x_2i + x_3 j + x_4 k\in \IH. \end{equation} 
We will consider connections the form
\begin{equation}A(r, t) = \frac{1}{2}\left\{ \frac{\bar{x}dx - d\bar{x} x}{f(r,t) + r^2}\right\} 
\qquad r = \sqrt{x_1^2 + x_2^2 + x_3^2 + x_4^2}.
\label{YMansatz}
\end{equation}
This is the most general form of a rotationally symmetric connection.  We will\
derive an equation of motion for $f(r,t)$ from
(\ref{genPDEa}).

To get at this, start with connections of the form
\begin{equation}A(r, t) = \frac{1}{2}g(r,t)\left\{\bar{x}dx - d\bar{x} x\right\}.\end{equation}
One can then use (\ref{genPDEa}) to calculate the differential equation for $g$:

\begin{equation} \partial_t^2{g} = 12 g^2 + \frac{5\partial_r g}{r} + \partial^2_r g - 8 g^3 r^2.\label{protPDEa}\end{equation}

Substituting $g(r,t) = (f(r,t) + r^2)^{-1},$ we obtain a differential equation for $f(r,t)$:
\begin{equation} \fddot = \drrf + \frac{5\drf}{r} - \frac{8 \drf r}{f + r^2} 
+ \frac{2}{f + r^2}\left((\fdot)^2 - (\drf )^2\right)
.\label{PDEa}\end{equation}

As with the charge-two sigma model, the static solutions are $f(r,t)=$
constant, where the constant is the square of the length scale, and
the geodesic approximation states that solutions should progress as
$f(r,t) = \alpha(t)$.  As before, blowup means $f(0,t)=0.$

\section{Numerical Method}

A finite difference method is used to compute the evolution of these partial 
differential equations (\ref{PDEb}, \ref{PDEc}, \ref{PDEa}) numerically.  
Centered differences are used consistently except for
\begin{equation} \drrf + \frac{3\drf}{r}\ \ \mbox{and}\ \  
\drrf + \frac{5\drf}{r} .\label{instabpart}\end{equation}
In order to avoid serious instabilities in 
(\ref{PDEb}, \ref{PDEc}, \ref{PDEa}) these are
modeled in a special way.  Let
\begin{equation} \mathcal{L}_1 f = 
r^{-3} (\partial_r) r^3 (\partial_r) f = \drrf + \frac{3\drf}{r}
\end{equation}
and 
\begin{equation} \mathcal{L}_2 f = r^{-5} (\partial_r) r^5 (\partial_r) f 
= \drrf + \frac{5\drf}{r}.
\end{equation}

These operators have negative real spectrum, hence they are stable.
The na\"{\i}ve central differencing scheme on (\ref{instabpart}) results in
unbounded growth at the origin, but the natural differencing scheme
for this operator does not.  For the first it is
\begin{equation} \mathcal{L}_1f \approx r^{-3}\left[ 
\frac{\left(r + \displaystyle{\frac{\delta}{2}}\right)^3
\left(\displaystyle{\frac{f(r+\delta) - f(r)}{\delta}}\right) 
- \left(r-\displaystyle{\frac{\delta}{2}}\right)^3
\left(\displaystyle{\frac{f(r) - f(r-\delta)}
{\delta}}\right)}{\delta}\right].\end{equation}
The second is done analogously.

With the differencing explained, we want to derive $f(r,t+\delt)$.  We
always have an initial guess at $f(r,t+\delt)$.  In the first time
step it is $f(r,t+\delt) = f(r,t) + v_0\delt$ with $v_0$ the initial
velocity given in the problem.  On subsequent time steps the initial guess is
$f(r,t+\delt) = 2f(r,t) - f(r,t-\delt)$.  The appropriate forward time step 
can be used to compute $\fdot(r,t)$
on the right hand side of (\ref{PDEb}, \ref{PDEc}, \ref{PDEa}).  
Then this allows us to
solve for a new $f(r,t+\delt)$ in the differencing for the second derivative
$\fddot(r,t)$.  This procedure can be iterated several times to get
increasingly accurate values of $f(r,t+\delt)$.  In (\ref{PDEa}), for example,
we iterate the equation:
\begin{eqnarray} f(r,t+\delt) &=& 2f(r,t) - f(r,t-\delt) + (\delt)^2\left[
\drrf(r,t) - \frac{5\drf (r,t)}{r}\right.\nonumber\\ & & - \left.
\frac{2(\partial_t{f}(r,t))^2}{f(r,t) + r^2} - \frac{2(\drf (r,t))^2}{f(r,t) + r^2}
- \frac{8 \drf (r,t)r}{f(r,t) + r^2}\right],\end{eqnarray} where all
derivatives on the right hand side are represented by the appropriate
differences.

There remains the question of boundary conditions.  At the origin,
in all models,
$f(r,t)$ is presumed to be a quadratic function, and this gives
\begin{equation} f(0,t) = \frac{4}{3} f(\delr,t) - \frac{1}{3} f(2\delr,t).\end{equation}
At the $r = R_{max}$ boundary we use Neumann boundary conditions
\linebreak[4] $\drf (R_{max},t) = 0$.  We essentially have a free wave
equation outside of this boundary in all of the equations.  This
boundary condition allows information that has propagated out to it to
bounce back toward the origin, but it does not permit information to
come in through the boundary from infinity.

In the models generated from (\ref{PDEc}, \ref{PDEa}), results indicate that 
the appropriate form for $f(r,t)$ at any given time $t$, is a parabola in $r$, rather than 
a straight line.  When looking at this aspect of the model, the $f(R_{max},t)$ boundary 
condition was changed to respect the parabola:
\begin{equation}\drf (R_{max},t) = \drf (R_{max}-\delr, t)\frac{R_{max}}{R_{max}-\delr}.\end{equation}  

\subsection{Stability and Accuracy}

A lengthy analysis of the stability of the numerical schemes for 
(\ref{PDEb}, \ref{PDEa}) is performed in
\cite{Linhart}.  
That for (\ref{PDEc}) should be directly analogous to that for
(\ref{PDEa}).  
We do not wish to go into the full details of the analysis for the stability
or convergence in this paper, as they are tedious.

However, we do emphasize the importance of such an analysis. 

Discrete numerical models, such as these, are usually approximations
to a desired continuum model.  Usually the true continuum solution is
only obtained as some parameter goes to zero or to infinity.  In
evaluating a model numerically, we introduce error.  We have the
unavoidable round off error that comes with every computer model, and
also a truncation error that occurs from only taking finitely many
terms.

Sometimes a method that seems otherwise favorable has undesired
behavior that is caused by the discretization used in the problem.

For example, questions arise from \cite{LPZ} about the stability of the lumps
for (\ref{PDEb}).  They found that a soliton would shrink without
perturbation from what should be the resting state.  This seems to be
a function of the numerical scheme used for those experiments.   The
numerical scheme used here for equations (\ref{PDEb}, \ref{PDEc}, \ref{PDEa})
has no such stability problems.  Experiments show that a stationary solution 
is indeed stationary unless perturbed by the addition of an initial velocity.

In unstable numerical schemes, the roundoff error can come to be
commingled with the calculation at an early stage and can then be
magnified until it overcomes the true answer.  A good, basic
discussion of stability and numerical schemes can be found in
\cite{NumRec}.

\begin{figure}[t]
\begin{center}
\epsfig{file=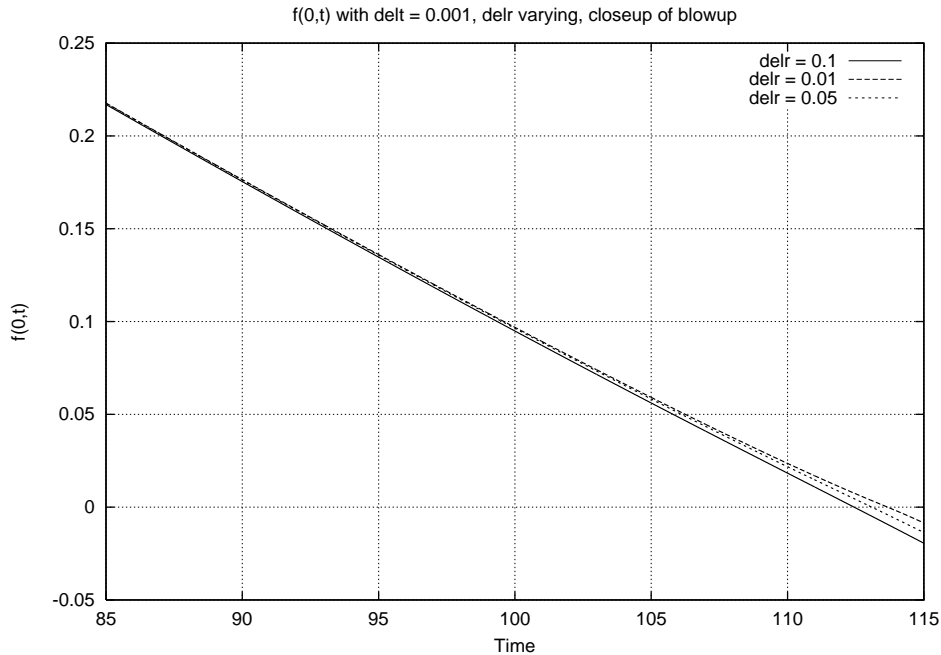}
\caption{Different trajectories
for different $\delr$ values in the charge-one $\sigma$-model.  
We have zoomed in on a neighborhood 
of the blowup time in order to
be able to see where these trajectories begin to deviate from one another.}
\label{conv2}
\end{center}
\end{figure} 

Apart from the stability analysis, one can measure the effect of the
spatial discretization by tracking different trajectories with
identical initial conditions, a small fixed $\delt$, and a varying
$\delr$.  Any scheme will break down when the system develops important
features that are too small compared to $\delr$. When two trajectories start
to differ, that implies that the computation with the larger value of $\delr$
has started to break down.

We see this in 
Figure \ref{conv2}, where we track $f(0,t)$ 
for the charge-one $S^2$ $\sigma$-model. 
We pick input
parameters of $f(r,0) = f_0 = 1.0$, $\partial_t f(r,0) = v_0 =
-0.01$, and $\delt = 0.001$.  We track evolution with $\delr = 0.1$,
$\delr = 0.05$ and $\delr = 0.01$.  The difference between
$f(0,t)_{\delr = 0.1}$ and $f(0,t)_{\delr = 0.01}$ at the point of
blowup (specifically, when $f(0,t)_{\delr = 0.01}$ hits zero) 
is less than $0.01$.  This
is considerably less than the grid size difference of $0.09$. Furthermore, 
the difference when $f(0,t)_{\delr = 0.1} = 0.1$ is smaller still,
with $|f(0,t)_{\delr = 0.1} - f(0,t)_{\delr = 0.01}| \approx 0.002$.
This correspondence in the three trajectories indicates that the
influence of the grid size on the features of the model can be ignored
until $f(0,t)$ has shrunk roughly to size $\delr$.

\section{Comparison to the Geodesic Approximation}

We wish to compare the actual evolution of the blowup of the charge-one 
and charge-two $S^2$ $\sigma$-models and the Yang-Mills model with 
that predicted by the geodesic approximation.

In our ansatz for all three models, we model the evolution of
$f(r,t)$, and in all cases the geodesics on the moduli space of
stationary solutions are $f(r,t) = \alpha(t)$.  When we calculate
deviations from the geodesic approximation of this profile, we
calculate deviations between the actual shape $f(r,T)$, with $T$
fixed, and $f(r,T) = \alpha$.  Note also that outside of a
sufficiently large ball about the origin, all three equations are well
approximated by the linear wave equation $\fddot = \drrf$.  The
interesting behavior should occur near the origin, and so evolution
toward blowup is tracked as the evolution of $f(0,t)$.

\subsection{Charge-One $S^2$ $\sigma$-Model}
For the charge-one $S^2$ $\sigma$-model, our calculations 
follow those done in \cite{PZ} and \cite{Speight1}. 

In \cite{PZ} the time evolution of blowup via the shrinking of lumps  
was studied.  The Lagrangian was cut off outside of a ball of
radius $R$, to prevent logarithmic divergence of the integral for the
kinetic energy, and then an analysis of what happens in the
$R\rightarrow \infinity$ limit was performed.  In \cite{Speight1} the problem of the
logarithmic divergence in the kinetic energy integral is solved by
investigating the model on the sphere $S^2$.  The radius of the sphere
determines a parameter for the size analogous to the parameter $R$ for
the size of the ball the Lagrangian is evaluated on in \cite{PZ}.

Equation (\ref{cp1lag}) gives us the Lagrangian for the general
version of this problem.  In the spherically symmetric 
geodesic approximation for the charge-one sector, we consider functions of
the form
\begin{equation} u = \frac{\beta}{z}.\end{equation}
If we restrict the Lagrangian to this space we get
an effective Lagrangian.  The integral of the spatial derivatives of
$u$ gives a constant, the Bogomol'nyi bound, and hence can be ignored.
If one integrates the kinetic term over the entire plane, one sees it
diverges logarithmically, so if $\beta$ is a function of time, the
solution has infinite energy.

Nonetheless, this is what we wish to investigate.  We presume
that the evolution takes place in a ball around the origin of size
$R$.  Up to a multiplicative constant, the
effective or cutoff Lagrangian becomes
\begin{equation} L = \int_0^R r dr \frac{r^2 (\fdot)^2}{(r^2 + f^2)^2} \end{equation}
which integrates to
\begin{equation} L = \frac{(\fdot)^2}{2} \left[ \ln\left(1 + \frac{R^2}{f^2}\right) -
\frac{R^2}{f^2 + R^2}\right].\end{equation} 
For geodesics, energy is conserved, so
\begin{equation} \frac{(\fdot)^2}{2} \left[ \ln\left(1 + \frac{R^2}{f^2}\right) -
\frac{R^2}{f^2 + R^2}\right] = \frac{c^2}{2}, \end{equation}
with $c$ (and hence $c^2/2$) a constant.
Solving for $\fdot$ we obtain
\begin{equation} \fdot = \frac{c}{\sqrt{\left[\displaystyle{\ln\left(1 + 
\frac{R^2}{f^2}\right) - \frac{R^2}{f^2 + R^2}}\right]}}.
\label{fdoteqn} \end{equation}
Since we are starting at some value $f_0$ and evolving toward the singularity at $f = 0$ this gives:
\begin{equation} \int_{f_0}^{f(0,t)} d\:\! f 
\sqrt{\ln\left(1 + \frac{R^2}{f^2}\right) - \frac{R^2}{f^2 + R^2}} =
\int_0^t c dt.\label{origevo}\end{equation} 

\begin{figure}[t]
\begin{center}
\epsfig{file=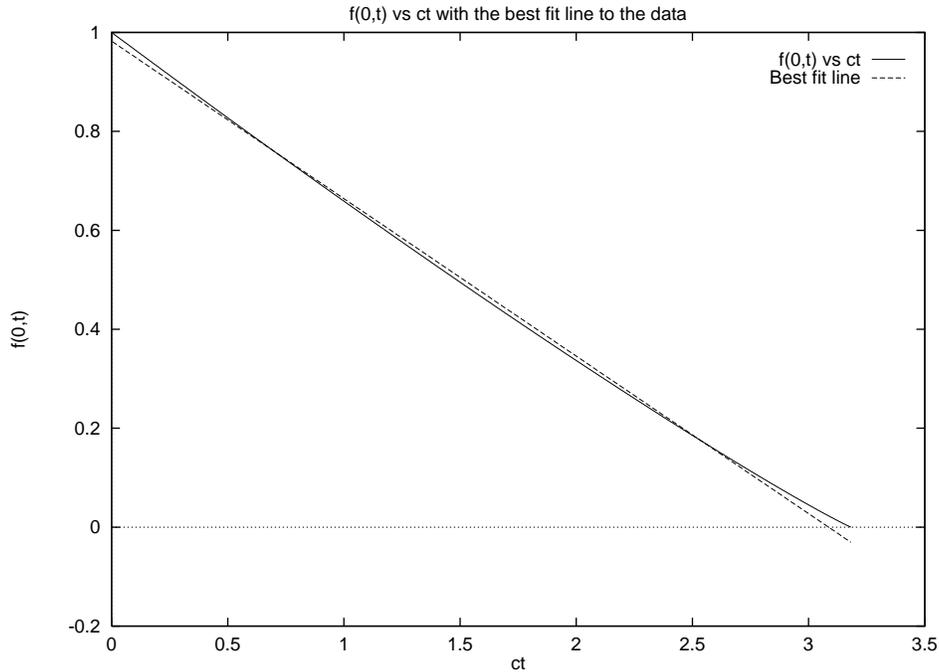}
\caption{Predicted plot of $f(0,t)$ vs $ct$ for the cutoff 
Lagrangian with $R=100$ in the charge-one $\sigma$-model.}
\label{exampc1}
\end{center}
\end{figure}

In the $R\rightarrow \infinity$ limit, as taken in \cite{PZ}, this
gives linear evolution, {\it i.e.,\/} fast blowup.  But this would
violate the Shatah-Struwe theorem \cite{SS}.  However, taking a finite value of
$R$ gives $\sqrt{|\ln(t)|}$ corrections and slow blowup, in accordance with
the theorem. 

In equation (\ref{origevo}), the integral on the right
gives $ct$.  The integral on the left can be evaluated numerically
for given values of $R$, $f_0$ and $f(0,t)$.  A plot can then be
generated for $ct$ vs. $f(0,t)$.  What we really are concerned with is
$f(0,t)$ vs. $t$, but once the value of $c$ is determined this can be
easily obtained.  One such plot with $f_0 = 1.0$, $R = 100$ of
$f(0,t)$ vs $ct$ is given in Figure \ref{exampc1}.  This curve is
not quite linear, as seen by comparison with the
best fit line; both are
plotted in Figure \ref{exampc1}.

\begin{figure}[t]
\begin{center}
\epsfig{file=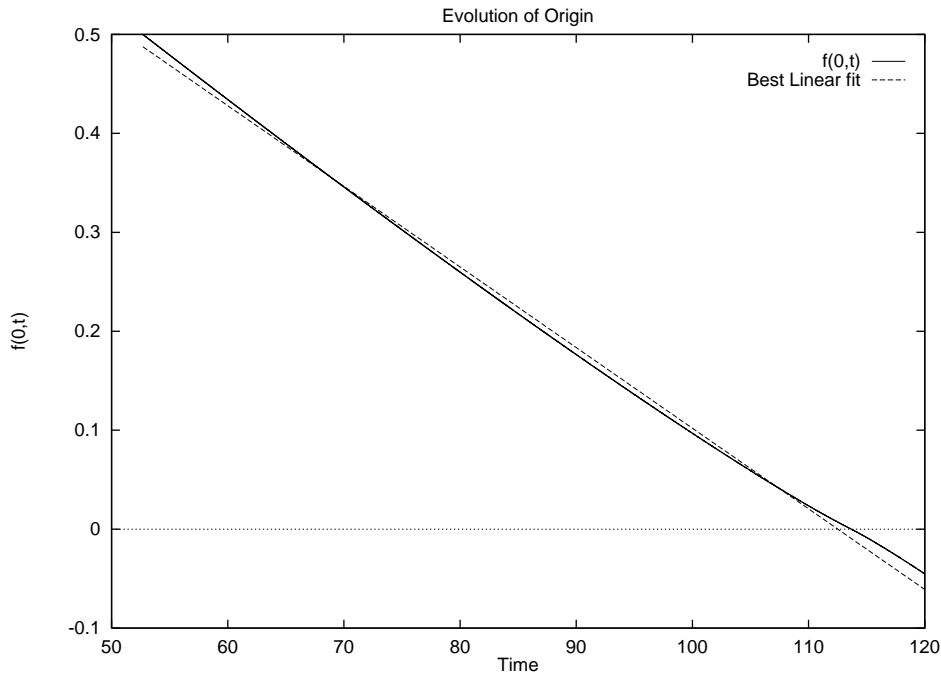}
\caption{Plot of $f(0,t)$ vs $t$ for an actual trajectory in 
the charge-one $\sigma$-model.  The slight nonlinearity is
reminiscent of the prediction shown in Figure \ref{exampc1}. }
\label{ocp1c1}
\end{center}
\end{figure}

The scale invariance of this problem allows us to consistently take
$f_0 = 1.0$ without loss of generality, and the key dimensionless
parameter is $\fdot(0,0)$.  We compare this with the numerical
computer model, which was run with various small velocities.  The
initial velocity is $\fdot(r,0) = v_0$.  Another important input
parameter is $R_{max}$, the largest value of $r$ modeled, which is
chosen large enough so that information from the origin cannot
propagate out to the boundary and bounce back to the origin within
the time to blowup.  Other input parameters are $\delr$ and $\delt$.

We track $f(0,t)$ as it heads toward this singularity, and
find that its trajectory is not quite linear, as seen in Figure
\ref{ocp1c1}.  This is suggestive of the result obtained in the
predictions for this model.

\begin{figure}[t]
\begin{center}
\epsfig{file=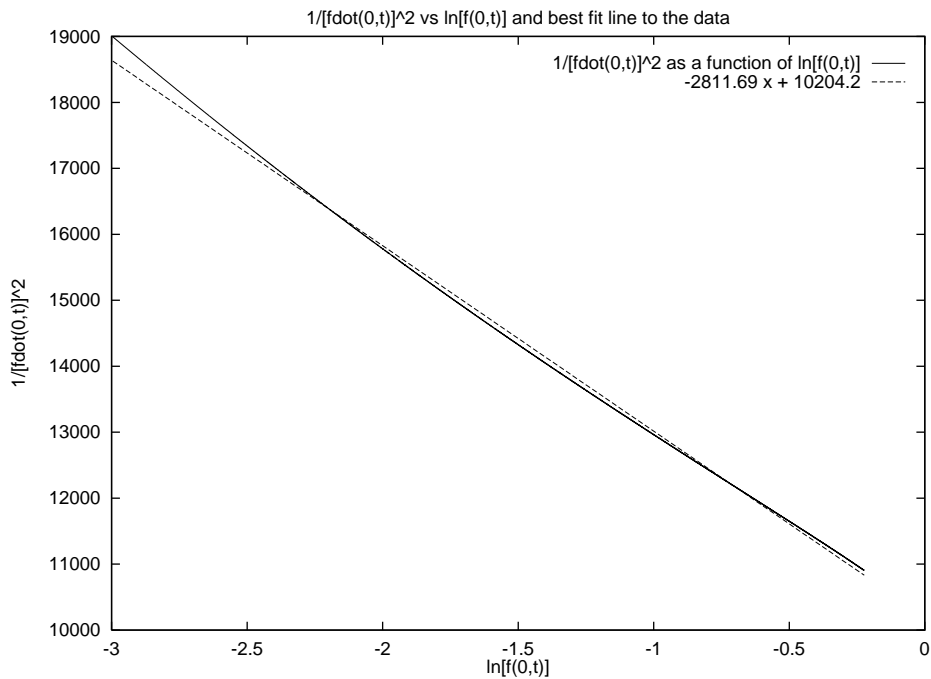}
\caption{Actual plot of $1/(\fdot(0,t))^2$ vs $\ln(f(0,t))$ for the 
charge-one $\sigma$-model and the best 
fit line to this data.  The cutoff Lagrangian predicts linear dependence, 
while the $R \to \infty$ limit predicts that $1/\fdot^2$ should be constant.}
\label{lnffdot}
\end{center}
\end{figure}

We want to check the legitimacy of the prediction (\ref{origevo}),
pictured in Figure \ref{exampc1}.  This requires a determination of
the cutoff\footnote{The Lagrangian cutoff $R$ is not to be confused
with $R_{max}$, the maximum value of $r$ modeled numerically.  $R$ is
a parameter of the geodesic approximation, not of the numerical
integration.} $R$ and the kinetic energy $c^2/2$.  We already have
$f_0$ and $f(0,t)$. To determine $R$ and $c$, observe from equation
(\ref{fdoteqn}) that:
\begin{equation} \frac{1}{(\fdot)^2} = \frac{\left[ 
\displaystyle{\ln\left(1 + \frac{R^2}{f^2}\right) - 
\frac{R^2}{f^2+R^2}}\right]}{c^2} \label{fdoteqnb}\end{equation}
Since $R$ is large and $f$ is small, we can rewrite 
equation (\ref{fdoteqnb}) as
\begin{equation} \frac{1}{(\fdot)^2} \approx \frac{\left[ 
\ln(R^2) - \ln(f^2) - 1\right]}{c^2}. \end{equation} 
The plot of
$\ln(f)= \ln(f(0,t))$ vs $1/{(\fdot)^2} = 1/{(\fdot(0,t))^2}$
should be linear with the slope $\ = 2/c^2$ and the intercept
$\ = (2\ln(R) - 1)/c^2$.  Such a plot is easily obtained from the
data, and the parameters $c$ and $R$ are
readily calculated.

\begin{figure}[t]
\begin{center}
\epsfig{file=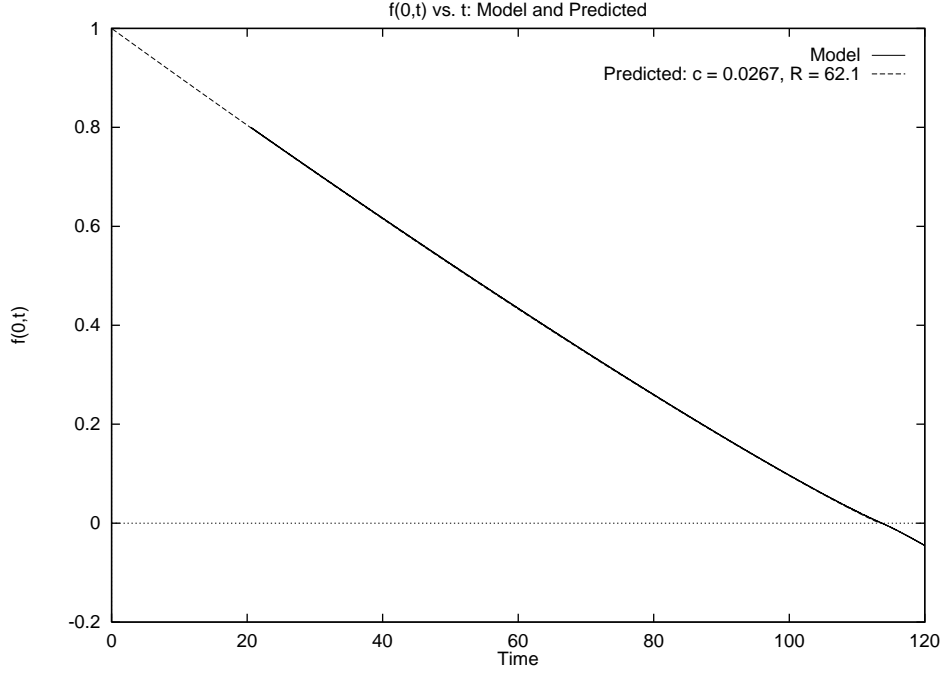}
\caption{Comparison of the cutoff geodesic approximation to actual data in 
the charge-one $\sigma$-model.
The predicted course of $f(0,t)$ vs. $t$ from equation 
(\ref{origevo}) is practically indistinguishable from the actual trajectory.}
\label{fevol}
\end{center}
\end{figure}

Figure \ref{lnffdot} is a plot of $\ln(f(0,t))$
vs. $1/(\fdot(0,t))^2$, with initial conditions $\delr = 0.01$, $\delt
= 0.001$, $f_0 = 1.0$ and $v_0 = -0.01$.  If the correct dynamics were
given by the $R\rightarrow \infinity$ limit of equation
(\ref{origevo}), this graph would be of a horizontal line, since the
velocity would be unchanging.  On the other hand, if equation (\ref{origevo}) is correct with $R$ finite, we should get a line with nonzero slope. 
It is easily seen that the plot of
$\ln(f(0,t))$ vs. $1/(\fdot(0,t))^2$ is nearly a linear relationship,
as predicted by the cutoff Lagrangian, but is not quite a straight line.
This may indicate that the effective values of $R$ and $c$ are
themselves changing slowly with time.

\begin{table}[t]
\begin{center}
\caption{ Best-fit parameters $c$ and $R$ as a function of $v_0$, with
$f_0 = 1.0$ for the charge-one $\sigma$-model.}
\label{crvtab}
\begin{equation} \begin{array}{{r}{l}{r}} v_0  & c & R \\
-0.005&   0.0145& 115\\
-0.00667& 0.0187&  89\\
-0.01&    0.0263&  53\\
-0.0133&  0.0342&  49\\
-0.02&    0.0485&  34\\
-0.03&    0.0683&  25\\
-0.05&    0.104&   17\\
-0.06&    0.121&   15 \\
\end{array}\end{equation}
\end{center}
\end{table}

The best fit line to this data has
slope -2810 and intercept 10200, corresponding to 
$c = 0.0267$ and $R = 62.1$.  Using these values of
$c$ and $R$ in the calculation of equation (\ref{origevo}), we obtain the plot
of $f(0,t)$ vs $t$ given in Figure \ref{fevol}.  This is overlaid
with the model data for $f(0,t)$ vs. $t$ for comparison.  These two
are virtually indistinguishable. 

We have seen that modifying the geodesic approximation by applying a
fixed and finite cutoff greatly improves its accuracy.  Moreover,
corrections to linear behavior are seen to be logarithmic, not
power-law.  However, we have not yet demonstrated any predictive
power, since the cutoff was deduced from the data only after the fact.
What is needed is an understanding of how the dynamically generated
cutoff $R$ (and the kinetic energy $c^2/2$) vary with initial
conditions.

\begin{figure}[t]
\begin{center}
\epsfig{file=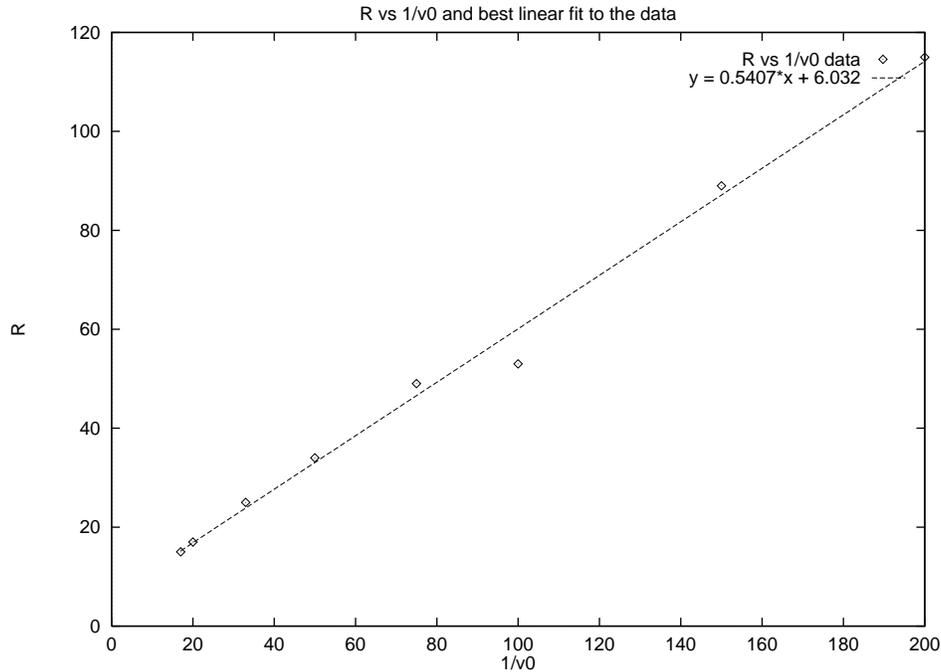}
\caption{The dynamical cutoff $R$  grows linearly with $1/v_0$ in the 
charge-one $\sigma$-model.}
\label{crvfit}
\end{center}
\end{figure}

Table \ref{crvtab} gives the best-fit values for $c$ and $R$ as a
function of the initial velocity $v_0$, under the initial conditions
$f_0 = 1.0$, $\delr = 0.01$ and $\delt = 0.001$.  $R$ is roughly linear in
$1/v_0$, with best-fit line
\begin{equation} R=  \frac{0.5407}{v_0} + 6.032.\end{equation}
This fit is shown in Figure {\ref{crvfit}}.  
A linear fit with $1/v_0$ makes sense, since the time to blowup is itself
of order $1/v_0$.  $R$ is roughly half the radius of the light cone at
$t=0$.

\begin{figure}[t]
\begin{center}
\epsfig{file=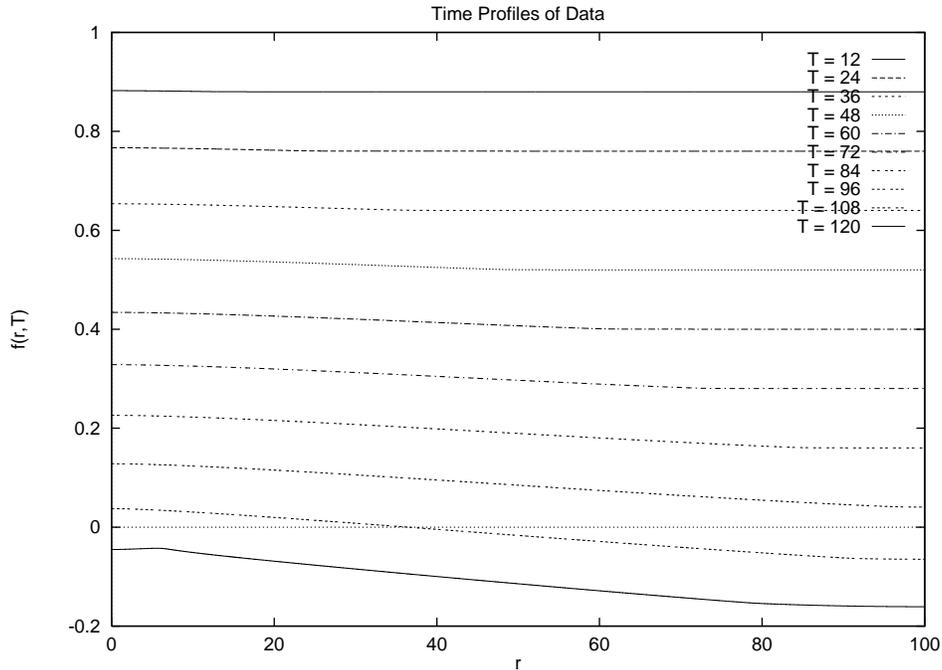}
\caption{Time slices $f(r,T)$ for the charge-one $\sigma$-model.}
\label{tscp1c1}
\end{center}
\end{figure}

\begin{figure}[t]
\begin{center}
\epsfig{file=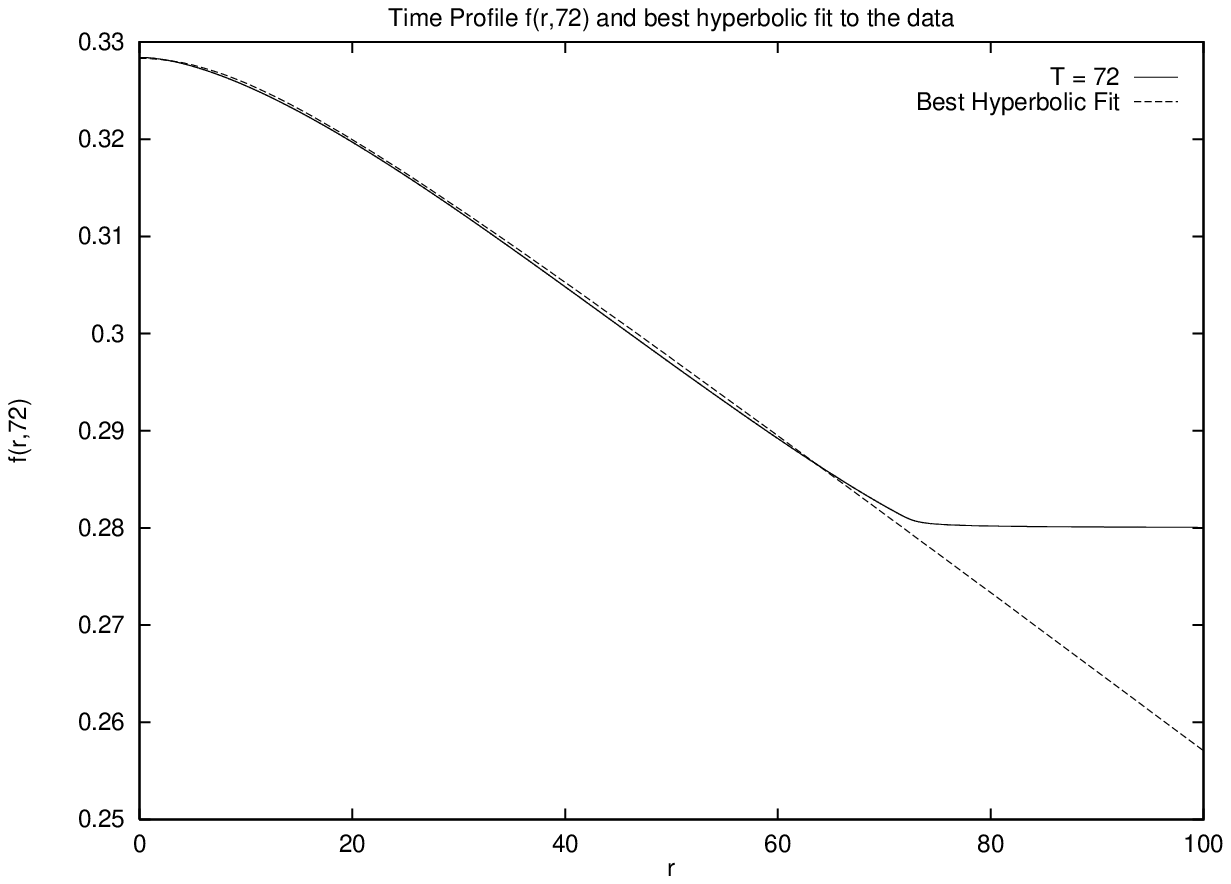}
\caption{Time slices for the charge-one $\sigma$-model 
evolve a hyperbolic bump at the origin.  
The transition from a hyperbola to a horizontal line
occurs at $r=T$.}
\label{hcp1c1}
\end{center}
\end{figure}

Next we consider the shape of the lump as we approach blowup.  
That is, what is $f(r,T)$ with $T$ fixed?
The geodesic approximation indicates that $f(r,t)$ should be independent
of $r$, but this is only approximately true. 
The graph of $f(r,t)$ versus $r$ 
stays close to horizontal, 
although there is some slope downward as time
increases.   This is shown in Figure \ref{tscp1c1}.

Making a closer inspection of the time profiles $f(r, T),$ 
as in Figure \ref{tscp1c1}, we observe that the initial part of
the data is close to a hyperbola as seen in Figure \ref{hcp1c1}.
This fit is close, 
but not nearly as close as those
for profiles in the charge-two sector of the $S^2$ $\sigma$-model or the
(4+1)-dimensional Yang-Mills Model, below.
While the asymptotic conditions for the hyperbola are not consistent 
with those of a solution to this equation, this form need only be 
assumed on a ball about the origin, outside of which the function is of the
form $f(r>R,T) = \alpha$.

\subsection{The Charge-Two $S^2$ $\sigma$-Model}

Equation (\ref{cp1lag}) gives us the Lagrangian for the general
version of this problem.  We are using
\begin{equation} u = \frac{\lambda(t)}{z^2}\end{equation}
for our evolution, and via the geodesic approximation we restrict the
Lagrangian integral to this space, to give an effective Lagrangian, as
we did in the previous section.  The integral of the spatial derivatives of $u$
gives a constant, and hence can be ignored.  Unlike the charge-one 
$\sigma$-model, this Lagrangian does not have an infrared
divergence, and there is no need to apply a cutoff.
Up to a multiplicative constant, the effective Lagrangian is
\begin{equation} L = \int_0^{\infinity} r dr \frac{r^4(\partial_t\lambda)^2}{\left(r^4 + \lambda^2\right)^2} = \frac{(\partial_t\lambda)^2\pi}{8\lambda}.
\end{equation}

By conservation of energy,
\begin{equation} \partial_t\lambda = 2 c_1 \sqrt{\lambda}, \end{equation}
for some constant $c_1$, and we integrate to obtain
\begin{equation} \lambda(t) = (c_1 t + c_2)^2,\end{equation}
which we rewrite as
\begin{equation} \lambda(t) = p(t-t_0)^2.\label{ch2predictor}\end{equation}
Equation (\ref{ch2predictor}) is how we predict that
$f(0,t)$ will evolve.

\begin{table}[t]
\begin{center}
\caption{Parameters for best fit parabola to $f(0,t)$ vs. initial data $f_0$ and $v_0$
for the charge-two $S^2$ $\sigma$-model.} 
\begin{equation}\begin{array}{*{3}{r}{l}}  f_0 &  v_0\ &\ \ \ t_0\ \ & p\ \\ 
1.0&    -0.01&   200&   0.0000250\\
1.0&    -0.02&   100&   0.0000998\\
1.0&    -0.03&    67&   0.000224\\
1.0&    -0.04&    50&   0.000398\\
1.0&     \frac{-\sqrt{2}} {100}  &   100\sqrt{2} &   0.0000499\\
1.0&    \frac{-0.01}{\sqrt{2}} &   200\sqrt{2} &   0.0000125\\
1.0&    \frac{-0.01}{\sqrt{3}} &   200\sqrt{3} &   0.00000833\\
\end{array} \nonumber\end{equation}
\label{fch2}
\end{center} 
\end{table}

\begin{figure}[t]
\begin{center}
\epsfig{file=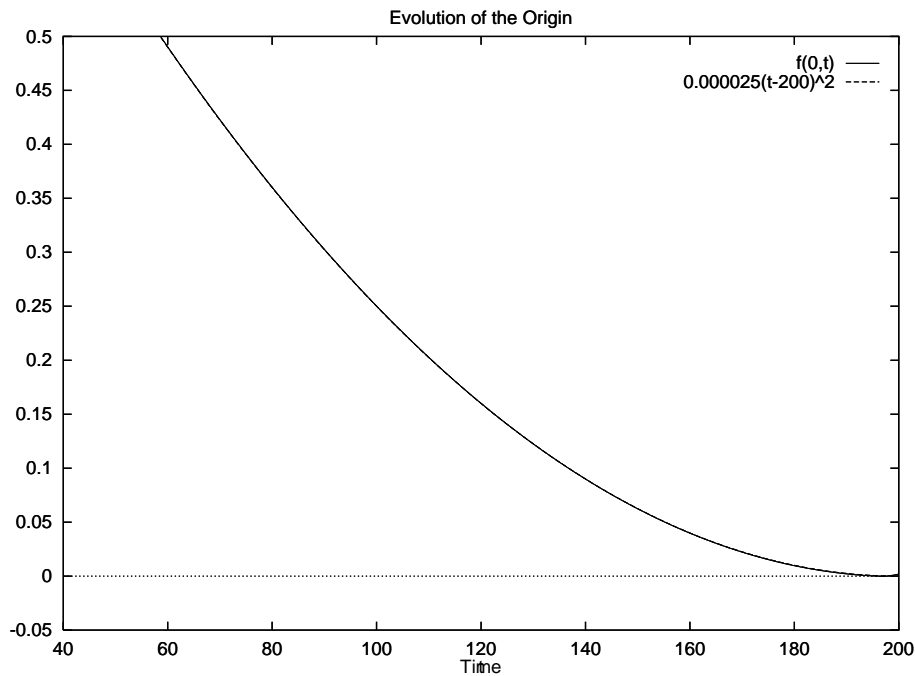}
\caption{Sample trajectory of $f(0,t)$ vs. $t$ in the 
Yang-Mills model.  The actual trajectory is almost
exactly parabolic.}
\label{o4p1}
\end{center}
\end{figure}

As before, the scale invariance of the model allows us to fix $f_0$,
and the other input parameters are the same as before.  A typical
evolution of $f(0,t)$ looks like that given in Figure
\ref{o4p1}. (That figure is actually a trajectory for the Yang-Mills
model, but the results from these two models are essentially the
same.)  In every case, the trajectory of $f(0,t)$ fits a parabola of 
the form $p(t-t_0)^2$ almost perfectly.
The dependence of the parabolic parameters $p$ and $t_0$ on
initial velocity $v_0$ is given in Table \ref{fch2}.  In all cases we took 
$f(r,0) = f_0=1.0$, $\partial_t{f}(r,0)=v_0$, 
$\delr = 0.025$ and $\delt = 0.001$, and in all cases we obtain 
\begin{equation} t_0 = \frac{2}{|v_0|}\qquad \hbox{and} \qquad
p = \frac{v_0^2}{4}.
\end{equation}
The predictability of the parameters from Table \ref{fch2}, and the 
closeness of the parabolic fits
indicate there is no correction needed to the predictions of the geodesic
approximation.  

With the evolution of $f(0,t)$ taken care of, we consider the shape of
the time slices $f(r,T)$ for a given fixed $T$.  The geodesic
approximation suggests that the graph of $f(r,T)$, which starts
horizontal, should remain horizontal.  Instead, an elliptical bump
forms at the origin, as seen in Figure \ref{tsl4p1}.  (That figure is
also from the Yang-Mills model, which exhibits essentially
identical behavior).

If we describe the elliptical bump by the equation
\begin{equation} \frac{x^2}{a^2} + \frac{(y-k)^2}{b^2} = 1,\label{ellbumpch2}
 \end{equation}
then the parameters $a$, $b$ and $k$ evolve as
\begin{eqnarray} a &=& t \\
 b &=& \frac{v_0^2}{4} t^2\\
 k &=& 1.0 + v_0t. 
\label{ellparms}\end{eqnarray}
The growth of this ellipse suggests that the curve is trying to adjust from
a horizontal line to a parabola
\begin{equation} f(r,t) = \rho r^2 + h, \label{parabch2}\end{equation}
where
\begin{equation}\rho =
-\frac{1}{2}\frac{d^2y}{dx^2} = 
-\frac{b}{2a^2}=-\frac{v_0^2}{8}.\label{rhoform}
\end{equation}  
To test this, we began  runs with initial data
\begin{equation}
f(r,0) = 1 - v_0^2 r^2/8; \qquad \partial_t f(r,0)=-v_0.
\end{equation}
for various values of $v_0$.
The time slices of these runs  retained a parabolic profile, as shown 
(for the Yang-Mills model) in Figure \ref{p4p1}.  That is, we expect
\begin{equation} f(r,t) \simeq -\frac{v_0^2}{8}r^2 
+ \frac{v_0^2}{4}\left(t - \frac{2}{|v_0|}\right)^2.\label{apxsol} 
\end{equation}
Substituting this closed-form expression 
into (\ref{PDEc}), we find the error is 
\begin{equation} \frac{v_0^6}{64}r^4 - \frac{v_0^6}{32}r^2\left(t - \frac{2}{|v_0|}\right)^2. \end{equation}
Within a fixed radius, and for small $v_0$, this error remains small for 
all time. 
However, the parabolic fit cannot be extended out arbitrarily far, especially
insofar as the profile (\ref{parabch2}) has logarithmically divergent
potential energy.

\subsection{The (4+1)-Dimensional Yang-Mills Model}

Equation (\ref{YMLagr}) gives us the Lagrangian for the general
version of this problem.  We are using
\begin{equation} A = \frac{1}{2} \left\{ \frac{\bar{x} dx - d\bar{x} x}{f + r^2}\right\}
\qquad r = \sqrt{x_1^2 + x_2^2 + x_3^2 + x_4^2},\end{equation} 
and we apply the
geodesic approximation. Restricting the Lagrangian to the moduli space
gives us an effective Lagrangian.  The portion of the integral given
by
\begin{equation}  -\frac{1}{4} \int_{\IR^4} |F_{ij}|^2 
d^4\vec{x} \end{equation}
represents the potential energy and integrates to a topological constant,
hence it may be ignored, as
in the study of the $S^2$ $\sigma$-model.  We need to calculate
\begin{equation} L = 
\frac{1}{2} \int_{\IR^4} |\partial_t{A_i}|^2 d^4\vec{x}. 
=\int_{\IR^4} \frac{3r^2 (\fdot)^2}{(f + r^2)^4} d^4\vec{x}
= \hbox{const.} \times \frac{(\partial_t f)^2}{f}. 
\end{equation}
Since energy is conserved, $\partial_t f$ is proportional to $f^{1/2}$,
and we integrate to get 
\begin{equation} f = p(t - t_0)^2.\end{equation}
In short, the predicted evolution is identical to that of the charge
two $\sigma$-model.

A typical evolution of $f(0,t)$ is given in Figure \ref{o4p1}.  In
this figure, the graph of $0.000025(t-200)^2$ neatly overlays the
graph of $f(0,t)$.  This picture represents the evolution where $f_0 =
1.0$ and $v_0 = -0.01$.

\begin{figure}[t]
\begin{center}
\epsfig{file=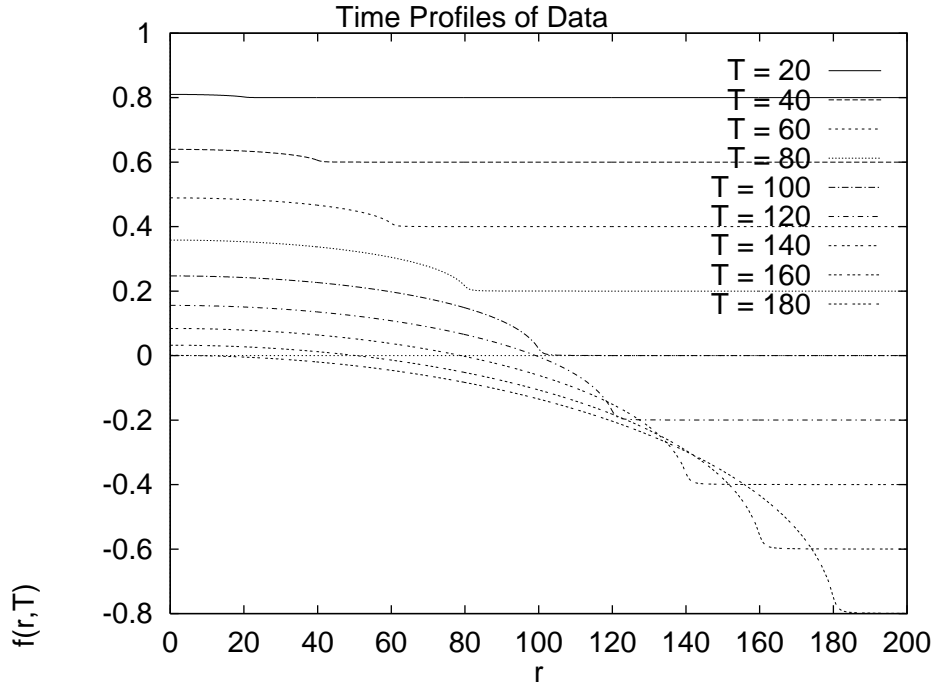}
\caption{Time slices $f(r,T)$ for the Yang-Mills model 
evolve an elliptical bump at the origin.  The edge of the bump is always 
at $r=T$.}
\label{tsl4p1}
\end{center}
\end{figure}

\begin{figure}[t]
\begin{center}
\epsfig{file=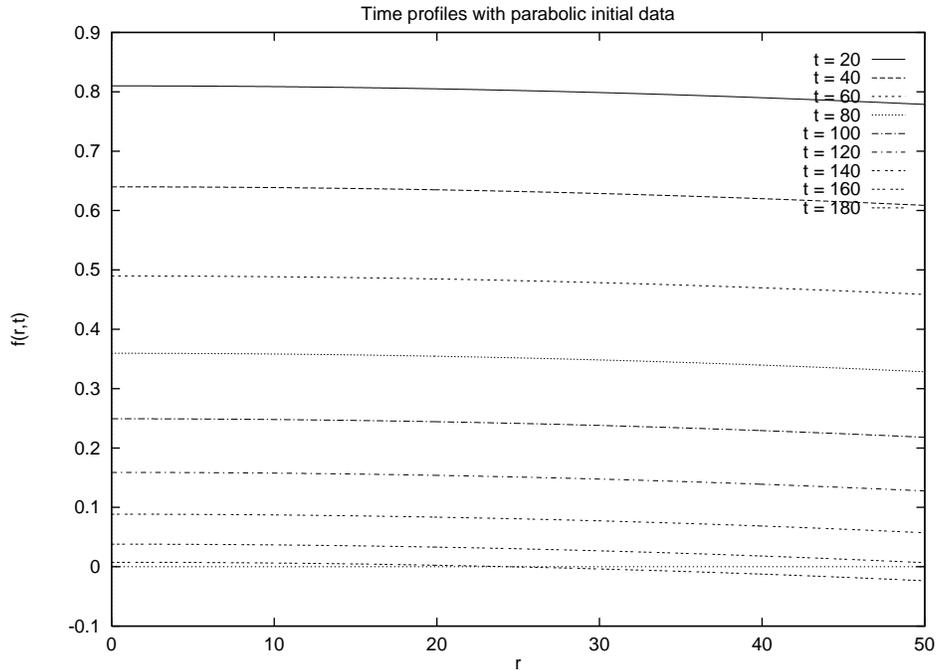}
\caption{With parabolic initial conditions,
the time slices in the (4+1)-dimensional Yang-Mills model do not form bumps, but instead remain parabolic.}
\label{p4p1}
\end{center}
\end{figure}

If we now look at the evolution of the initial line, $f(r,0) =
f_0,$ 
we obtain the same striking result as with the charge-two $S^2$ $\sigma$-model.
The initial line, $f(r,0) =
f_0,$ evolves an elliptical bump at the origin that grows as time
passes.   Figure \ref{tsl4p1} shows this behavior.
The clear indication here is that corrections to the profile are quadratic.
The elliptical bumps can be modeled exactly as before. 
Also as before, this suggests that we look for an approximate solution 
of the form $f(r,t)=\rho(t) r^2 +h(t)$.
Calculating from the general form of our ellipse
in (\ref{ellbumpch2}) we get the same results for $\rho$ as in (\ref{rhoform}),
and $h$ should evolve the same way as $f(0,t)$.  
%
When a run is started with this initial data $\partial_t{f_0} = v_0 =
-0.01$, $f(r,0) = 1.0 - \rho r^2$ and $\rho = -\frac{v_0^2}{8} =
-0.0000125$, the time slices of the data retain a parabolic profile.  This
is shown in Figure \ref{p4p1}.
The curvature of the parabola at the origin, as measured by the
parameter $\rho$ from equation (\ref{parabch2}) changes by less than 1
part in 100 during the course of this evolution.

The general
form of a parabolic $f(r,t)$ is the same as in equation (\ref{apxsol}).
If we substitute this into the partial differential equation (\ref{PDEa}),
we get an error of
\begin{equation} 2\left(\frac{v_0^2}{4}\right)^2r^2.\end{equation}
Since our concern is the adiabatic limit, $v_0$ is always
chosen to be small, and this difference goes to zero quickly in the adiabatic
limit.  

The form of the connection $A$ given by this parabolic model is
\begin{equation} A(r,t) = \frac{1}{2} \left\{ \frac{\bar{x}dx - d\bar{x}x}
{\left(1 -\frac{v_0^2}{8}\right)r^2 + \frac{v_0^2}{4} \left(t -
\frac{2}{|v_0|}\right)^2}\right\}.\end{equation} This is close to the
connection $A$ given by the geodesic approximation multiplied by an
overall factor. This form  suggests a relativistic correction.

\section{Conclusions}

We have studied the time evolution of solutions towards a singularity 
in the charge-one $S^2$ $\sigma$-model, the charge-two 
$S^2$ $\sigma$-model and the (4+1)-dimensional Yang-Mills model.  
We use a simple and robust 
numerical scheme to do so, for which the stability has been verified.  
The importance of the stability analysis is that we have no 
problem trying to separate real effects from numerical 
artifacts.

In the charge-one $\sigma$-model, the geodesic approximation is 
ill-defined due to the divergence of the Lagrangian.  A theorem of 
\cite{SS} tells us that there is no fast blowup in this
model.  The divergence of the geodesic approximation is 
solved by cutting off the Lagrangian at a radius $R< \infinity$.  
In the $R \rightarrow \infinity$ limit of this cut-off
Lagrangian, we have a prediction for linear evolution, 
which would be fast blowup in violation of 
the theorem in  \cite{SS}.  However, the cut-off Lagrangian  
itself is extremely accurate in predicting the shrinking of lumps 
toward the singularity
in this model.  This evolution is close 
to linear with complicated logarithmic corrections (equation (\ref{origevo})) 
going approximately as $\sqrt{\ln(t_0-t)}$.  Corrections from linear are not 
power-law, nor power-law 
corrected by a logarithm, as suggested in \cite{PZ}.  

We model the charge-two $\sigma$-model and the (4+1)-dimensional
Yang-Mills model to provide a comparison to our charge-one $S^2$
$\sigma$-model results.  We verify the belief that the charge two
$S^2$ $\sigma$-model is similar to the (4+1)-dimensional Yang-Mills
model; the numerical results show that the two models are almost
identical in behavior.  There is no theorem restricting fast blowup in
either of these models, and we find fast blowup as predicted by the
geodesic approximation in both.  The numerical models' evolution
toward the singularity is in lock step with prediction, and no
corrections are needed to this order.  Since the stability analysis of
the numerical schemes in the charge-one $S^2$ $\sigma$-model and the
Yang-Mills model is similar, the accuracy of the numerics to
prediction in the charge-two $S^2$ $\sigma$-model and Yang-Mills model
provide a accuracy scale on which to judge the charge-one $S^2$
$\sigma$-model.

In addition to modeling the evolution toward the singularity, we also can
characterize the shape of the profiles of the lumps with the time
variable fixed.  In the geodesic approximation, with the
rotationally symmetric functions we have chosen to model, 
the predicted evolution in all three cases was $f(r,t) = \alpha(t)$.  
The actual profiles in all three cases
have quadratic corrections.  

\medskip

\par
\noindent{\bf Acknowledgments}

\medskip

We thank Martin Speight for some extremely helpful suggestions.  This
work was partially supported by the Texas Advanced Research Program.

\pagebreak[4]

\bibliography{art}
\end{document}